\documentclass[aps,amsfonts,prd,preprint,nofootinbib]{revtex4}
\usepackage{graphicx}
\def\thebibliography#1{\leftline{\large\bf References}\list
  {[\arabic{enumi}]}{\settowidth\labelwidth{[#1]}\leftmargin\labelwidth
\advance\leftmargin\labelsep
\usecounter{enumi}}
\def\newblock{\hskip .11em plus .33em minus .07em}
\sloppy\clubpenalty4000\widowpenalty4000}

\newcommand{\Tr}{\mbox{Tr}}
\newcommand{\tr}{\mbox{tr}}
\newcommand{\fourier}[1]{\tilde{#1}}
\newcommand{\classicalVacuum}{zero-classical-energy }
\newcommand{\ClassicalVacuum}{Zero-classical-energy }
\newcommand{\HiggsGauge}{Higgs-Gauge }
\newcommand{\oneF}{_{\rm eff}^{(1)}}
\newcommand{\zeroF}{_{\rm eff}^{(0)}}
\newcommand{\zr}[1]{\mbox{\hspace*{#1em}}}
\newcommand{\ID}{\mbox{{\sf 1}\zr{-0.16}\rule{0.04em}{1.55ex}\zr{0.1}}}
\begin{document}

\begin{center}
{\large \bf Heavy Fermion Quantum Effects in $SU(2)_L$ Gauge Theory }
\\[1em] 
E.~Farhi$^{\rm a}$, N.~Graham$^{\rm b}$,
R.L.~Jaffe$^{\rm a}$, V.~Khemani$^{\rm a}$,
H.~Weigel\footnote{Heisenberg Fellow}$^{\rm a,c}$
\\[2em]

$^{\rm a}$Center for Theoretical Physics,
Laboratory for Nuclear Science\\ and Department of Physics,
Massachusetts Institute of Technology\\ Cambridge, Massachusetts 02139 \\[1em]
$^{\rm b}$Department of Physics, Middlebury College\\ 
Middlebury, VT  05753 \\[1em]
$^{\rm c}$Institute for Theoretical Physics, T\"ubingen University\\
D-72076 T\"ubingen, Germany\\[2em]

{\qquad \rm \qquad MIT-CTP-3350 \qquad UNITU-HEP-4/2003 }

\end{center}
\bigskip
\centerline{\large\bf Abstract}

We explore the effects of a heavy fermion doublet in a simplified version of the standard electroweak theory.   We integrate out the doublet and compute the exact effective energy functional of spatially varying gauge and Higgs fields.  We perform  a variational search  for a local minimum of the effective energy and do not find evidence for a soliton carrying the quantum numbers of the decoupled fermion doublet.   The fermion vacuum polarization energy offsets the gain in binding energy previously argued to be sufficient to stabilize a fermionic soliton.  The existence of such a soliton would have been a natural way to maintain anomaly cancellation at the level of the states.   We also see that the sphaleron energy is significantly increased due to the quantum corrections of the heavy doublet.  We find that when the doublet is slightly heavier than the quantum--corrected sphaleron, its decay is exponentially suppressed owing to a new barrier.  This barrier exists only for an intermediate range of fermion masses, and a heavy enough doublet is indeed unstable. 

\newpage
\section{Introduction}

In a chiral gauge theory like the Standard Model, gauge invariance
prevents fermions from having explicit mass terms.  Rather, they
get their mass through their coupling to a scalar field via the
well-known Higgs mechanism.  At tree level the fermion obtains
a perturbative mass, the product of the corresponding 
Yukawa coupling and the vacuum expectation value of the scalar Higgs 
field.  The decoupling of such fermions presents interesting unsolved 
puzzles.  Ordinary decoupling arguments \cite{Appelquist:tg}, which would show that a heavy
fermion is irrelevant in the low-energy theory, break down.
Increasing the mass, which causes the denominators in the fermion
propagators to suppress quantum corrections, also increases the coupling,
which gives a corresponding enhancement from the vertices.
Furthermore, unlike an ordinary fermion with vector couplings, a
chiral fermion cannot simply disappear from the theory as its mass is
increased, because anomaly cancellation would be ruined.  As shown in
ref.~\cite{DHokerFarhi}, gauge invariance is maintained at the level
of the Lagrangian because integrating out the heavy fermion induces a
Wess-Zumino term in the resulting effective Lagrangian.

For the case of Witten's non-perturbative $SU(2)$ anomaly \cite{Witten},
one can analyze the theory at the level of the action along the same
lines \cite{DHokerFarhi}.  However, one can also analyze the situation from
a different point of view.  Ref. \cite{FarhiSkyrme} 
shows that this anomaly can be understood in terms of the Hamiltonian of the 
theory:  A theory with an odd number of left-handed fermion doublets
has no gauge-invariant states.  However, if the Yukawa coupling 
of a fermion is large enough, the perturbative fermion mass will be
larger than the classical energy of the sphaleron \cite{Manton}, so that such
fermions are no longer stable states in the spectrum of the theory.
Thus, to maintain gauge invariance in the low-energy theory, there
must exist either new states in the theory carrying the quantum
numbers of the fermion, or a mechanism to suppress the decay of the
perturbative fermion states.

Although these scenarios could rely on complicated non-perturbative
physics, one simple resolution would be provided by the existence of a
soliton carrying the quantum numbers of the decoupled fermion.
If a localized configuration of gauge and Higgs fields binds a fermion
level tightly, the binding energy could outweigh the cost in classical
energy  to set up the background field configuration.  However, to
consistently include the effects of the fermion level, such
calculations must also include the Casimir energy, the renormalized
shift in the zero-point energies of all the other fermion modes, since
both appear at the same order in $\hbar$.  For a static field configuration, the Casimir energy
represents the full one-loop quantum vacuum polarization energy, equivalent to summing to all orders in the
derivative expansion.

To compare the quantum energy of the configuration with
the sphaleron, we must also include the corresponding correction to
the sphaleron's energy as well.  Furthermore, we must check whether
the quantum corrections induce an energy barrier between the 
perturbative fermion and the sphaleron, which would mean that the
perturbative fermion would be quasi-stable, only able to decay by tunneling.

We have carried out such calculations in a simplified version of the
Standard Model, for background fields in the spherical ansatz, keeping fermion vacuum fluctuations but ignoring those of the gauge and Higgs fields.  We find significant quantum corrections to the height of the sphaleron barrier.  As we
make the fermion level heavier than the quantum corrected sphaleron, we do see evidence for a barrier
suppressing its decay. For even larger Yukawa couplings, however, the
barrier disappears and the fermion's decay is unsuppressed.   We do
not see any evidence for a soliton for any value of the Yukawa
coupling, and find that including the full Casimir energy destabilizes
solitons found in previous work \cite{NolteKunz}.

We comment on possible explanations of this result and
its relation to the Witten anomaly in the conclusions.

\section{The Theory}
We consider the electroweak sector of the Standard Model with three
simplifying modifications: (1) the $U(1)$ hypercharge gauge fields are
absent, (2) the fermions within an isospin doublet are degenerate in
mass, and (3) the Cabibbo-Kobayashi-Maskawa (CKM) matrix is the
identity, so there is no mixing between fermions of different
generations.  Since there is no $U(1)$, the theory does not have
any perturbative gauge anomalies, but Witten's non-perturbative
anomaly requires  it to have an even number of $SU(2)_L$
fermion doublets.

The \HiggsGauge sector Lagrangian density is
\begin{equation}
\mathcal{L}_H =  -\frac{1}{2} \tr \left(F^{\mu\nu}F_{\mu\nu}\right) 
+ \frac{1}{2}\tr 
\left(\left[D^{\mu}\Phi \right]^{\dag} D_{\mu}\Phi\right) - 
\frac{\lambda}{4}\left[ 
\tr\left(\Phi^{\dag}\Phi\right) - 2v^2 \right]^2 \, ,
\label{HiggsLagrangian}
\end{equation}
where
\begin{eqnarray}
F_{\mu\nu} & = & \partial_\mu A_\nu - \partial_\nu A_\mu - i g\left[ A_\mu 
, A_\nu \right] \, , \nonumber \\
D_\mu \Phi & = & \left( \partial_\mu - i g A_\mu \right)\Phi \, , \nonumber 
\\
A_\mu & = & A_{\mu}^a \frac{\tau^a}{2} \, , 
\label{covariant}
\end{eqnarray}  
and where $g$ and $\lambda$ are gauge and Higgs 
self--interaction coupling constants respectively. Furthermore,
$v$ denotes the tree--level vacuum expectation value of the 
Higgs field. The $2 \times 2$ matrix field $\Phi$ is related to the Higgs doublet 
$\phi$ by 
\begin{equation}
\Phi =  \left( \begin{array}{cc} \phi_2^{*} & \phi_1 \\ -\phi_1^{*} & 
\phi_2 \end{array} \right) \, .
\end{equation}
$\Phi$ can be re-expressed in terms of four real functions as 
\begin{equation}
\Phi(x) = v\left( s(x) + i p^a (x) \tau^a \right)\, .
\end{equation} 
The fermionic sector Lagrangian density for each isospin doublet is
\begin{equation}
\mathcal{L}_F = \overline{\Psi}_L i\gamma^\mu D_\mu \Psi_L + 
\overline{\Psi}_R i\gamma^\mu \partial_\mu \Psi_R - f\left( 
\overline{\Psi}_L \Phi\Psi_R + \mbox{h.c.} \right) \, ,
\end{equation}
where the covariant derivative is the same as in eq.~(\ref{covariant})
and $f$ is the Yukawa coupling constant (which may be different for each doublet).  
There is no coupling between fermions belonging to different isospin 
doublets owing to our assumption of diagonal CKM matrix.  We introduce the
potential
\begin{equation}
V(A, \Phi) = -g\gamma^\mu A_\mu (x) \frac{1-\gamma_5}{2} + f\left(h(x)
+ i v p^a (x) \tau^a\gamma_5 \right) \, ,
\label{potential}
\end{equation}
where
\begin{equation}
h(x) \equiv v(s(x)-1) \, ,
\end{equation}
to write this Lagrangian density as the sum of a free part and an interaction part:
\begin{equation}
\mathcal{L}_F = \overline{\Psi} \left( i\gamma^\mu \partial_\mu - fv\right)\Psi - \overline{\Psi} V \Psi \, .
\label{Lf}
\end{equation}
We are interested in decoupling a single fermion doublet from the 
theory and for the remainder of the paper we consider only this single 
doublet.  The full theory is defined by
\begin{equation}
\mathcal{L} = \mathcal{L}_H + \mathcal{L}_F \, .
\end{equation}

In the unitary gauge ($p^a(x)=0$) at tree level, we have a 
single Higgs particle, $h(x)$, with perturbative mass $m_h^{(0)} = 2v\sqrt{\lambda}$. The superscript `(0)' denotes that the mass is at tree level.  The three gauge fields are degenerate with perturbative mass $m_w^{(0)}=g
v/\sqrt{2}$.  The two degenerate fermions have perturbative mass $m_f^{(0)}=f v$.
\section{The \HiggsGauge Sector Effective Energy}
We integrate out the fermion doublet from the theory to obtain the 
effective action for the \HiggsGauge sector:
\begin{equation}
e^{i S_{\rm eff}[A,\Phi]} = e^{i\int d^4 x {\mathcal L}_H} \frac{
\int[d\Psi][d\overline{\Psi}] e^{i\int d^4x {\mathcal L}_F}}{ \int[d\Psi][d\overline{\Psi}] e^{i\int d^4x {\mathcal L}_F |_{\rm V=0}}} \, .
\end{equation}
The normalization has been chosen so that the effective action is equal to the classical action for vanishing interaction potential, $V$, defined in eq.~(\ref{potential}).
If $i S_{\rm FD}^{(n)}$ denotes the Feynman diagram with one fermion loop and 
$n$ external insertions of $\left(-i V(A,\Phi)\right)$, then
\begin{equation}
S_{\rm eff}[A, \Phi] = \int d^4 x {\mathcal L}_H + \sum_{n=1}^{\infty} 
S_{\rm FD}^{(n)}\, .
\label{SeffFD}
\end{equation}
The Feynman diagrams can be computed in a prescribed 
regularization scheme. The divergences that emerge as the
regulator is removed are canceled by counterterms. We introduce
the renormalized and counterterm Lagrangians, ${\mathcal L}_H^{\rm (ren)}$ and ${\mathcal L}_H^{\rm (ct)}$ respectively,
by expressing the bare parameters in ${\mathcal L}_H$ in terms of 
renormalized parameters and counterterm coefficients to obtain ${\mathcal L}_H = 
{\mathcal L}_H^{\rm (ren)} + {\mathcal L}_H^{\rm (ct)}$.   The renormalized Lagrangian is  the original \HiggsGauge sector Lagrangian, eq.~(\ref{HiggsLagrangian}), with
renormalized parameters substituted and 
\begin{eqnarray}
{\mathcal L}_H^{\rm (ct)} & = & 
c_1\tr\left(F^{\mu\nu}F_{\mu\nu}\right) + 
c_2\tr\left(\left[D^{\mu}\Phi\right]^{\dag}D_{\mu}\Phi\right )
\nonumber \\ & & +\ 
c_3\left[\tr\left(\Phi^{\dag}\Phi\right)-2v^2 \right] + 
c_4\left[\tr\left(\Phi^{\dag}\Phi\right)-2v^2\right]^2 \, .
\label{Lct}
\end{eqnarray}
The coefficients $c_i$ depend on the regulator.   
 For notational simplicity we do not introduce a different notation for the renormalized parameters. 

If we consider static \HiggsGauge fields and restrict time to the 
interval $T$, then the effective energy functional is
\begin{equation}
E_{\rm eff}[A,\Phi] =  
- \lim_{T \rightarrow \infty} \frac{1}{T}\, S_{\rm eff}[A,\Phi]  
\,\equiv\,  E_{\rm cl}[A,\Phi] + E_{\rm vac}[A,\Phi] \, ,
\label{Eeff}
\end{equation}
where $E_{\rm cl}$ refers to the classical energy of the \HiggsGauge
sector.  We use the freedom to make time-dependent gauge
transformations to set $A_0=0$, so that
\begin{equation}
E_{\rm cl}[A,\Phi] =  
\int d^3x \left\{\frac{1}{2} \tr\left(F_{i j}F_{i j}\right) + 
\frac{1}{2}\tr\left(\left[D_{i}\Phi\right]^{\dag} D_{i}\Phi\right) + 
\frac{\lambda}{4}\left[\tr\left(\Phi^{\dag}\Phi\right)-2v^2\right]^2  
\right\} \, . 
\label{Eclassical}
\end{equation}
The fermionic vacuum polarization energy is
\begin{equation} 
E_{\rm vac}[A,\Phi] =  
\sum_{n=1}^{\infty} E_{\rm FD}^{(n)}[A,\Phi] + E_{\rm ct}[A,\Phi]\, ,
\label{Evac1}
\end{equation}
where each regulated Feynman diagram contribution is
\begin{equation}
E_{\rm FD}^{(n)}[A,\Phi]  =  
- \lim_{T \rightarrow \infty} \frac{1}{T}S_{\rm FD}^{(n)}\, , 
\end{equation}
and the counterterm contribution is
\begin{eqnarray}
E_{\rm ct}[A,\Phi]  & = & \int d^3x \Biggl\{ 
-c_1\tr\left(F_{i j}F_{i j}\right) 
+c_2\tr\left(\left[D_{i}\Phi\right]^{\dag} D_{i}\Phi\right)
\nonumber \\ & & \hspace{1.5cm}
-c_3\left[\tr\left(\Phi^{\dag}\Phi\right)-2v^2\right] 
-c_4\left[\tr\left(\Phi^{\dag}\Phi\right)-2v^2\right]^2 \Biggr\} \, .
\end{eqnarray}
The entire one--loop effective energy receives contributions also from gauge and Higgs loops.  We ignore these contributions because we believe the fermion loops are fundamentally responsible for the phenomena associated with fermion decoupling.  If we imagine that the fermions have $N_C$ internal degrees of freedom (e.g. color), then this approximation becomes exact for large $N_C$.  Nevertheless we set $N_C=1$ in the analysis that follows.
\subsection{The Counterterms}
The counterterms render $E_{\rm vac}$ finite by canceling the divergences in $E_{\rm FD}^{(n)}$, for $n=1$
through $n=4$, that arise when the regulator is removed. To unambiguously determine the 
counterterm coefficients, $c_i$, we impose
conventional renormalization conditions:
\begin{itemize}
\item[a.]
We choose the vacuum expectation value (vev) 
of $h(x)$ to be 0, which
ensures that the vev $\langle \Phi \rangle = v\ID$ stays fixed at its classical value and
the perturbative fermion mass does not get 
renormalized, {\it i.e.} $m_f = m_f^{(0)}$.  This is equivalent to a ``no--tadpole''
renormalization condition and determines $c_3$.

\item[b.]
We fix the pole of the Higgs propagator to be at
the tree level mass, $m_h=m_h^{(0)}$, 
with residue 1. These conditions yield
$c_2$ and $c_4$.

\item[c.]
We have various choices to fix the remaining undetermined 
counterterm coefficient $c_1$. We choose to set the
residue of the pole of the gauge field propagator to 1 
in unitary gauge. Then the position of that pole, {\it i.e.} the mass
of the gauge field, $m_w$, is a prediction.
\end{itemize}
The resulting counterterm coefficients, $c_i$, are listed
in Appendix A.
As explained under item c., the mass of gauge fields 
is constrained by the other model parameters when fermion 
loops are included. With our choice of renormalization conditions,
it is the solution to the implicit equation
\begin{eqnarray}
m_w^2 & = & \left(m_w^{(0)}\right)^2
\Biggl[1+\frac{f^2}{8\pi^2} 
\Biggl\{\frac{2}{3}-\frac{m_w^2}{m_f^2}
\left(\frac{1}{6}-\int_0^1dx x^2(1-x)^2\frac{m_w^2}
{\Delta(x,m_w^2)}\right) 
\nonumber \\ && \hspace{3cm}
+6\int_0^1 dx x(1-x)\ln\frac{\Delta(x,m_h^2)}{m_f^2} 
-\int_0^1dx\ln\frac{\Delta(x,m_w^2)}{m_f^2} \Biggr\} \Biggr] \, ,
\label{ModelParamsConstraint}
\end{eqnarray}
with $\Delta (x, q^2) \equiv m_f^2 - x(1-x)q^2$.  Recall that $m_w^{(0)} = gv/\sqrt{2}$ is the tree level perturbative mass of the gauge fields.
\subsection{The Vacuum Polarization Energy}
We briefly summarize methods introduced in refs.
\cite{PhaseshiftsGeneral,PhaseshiftsFermions}
(see ref.~\cite{Leipzig} for a review and a list of 
   additional references) that enable an {\em
exact} calculation of the fermionic vacuum polarization energy.  
Eq.~(\ref{Evac1}) gives this energy as an infinite sum of Feynman
diagrams.  For non--perturbative field configurations
all orders would have to be summed, which is intractable. 
We therefore make use of the fact that it is also given by a sum over the shift in the zero-point
energies of the fermion modes due to the background fields.  We write this
formal quantity as a sum over bound state energies, $\epsilon_j$,
(times their degeneracies, $D_j$)
and a momentum integral of the energy 
of the continuum states weighted by the change in the density 
of continuum states, $\Delta\rho(k)$, that is
induced by the background fields,
\begin{equation}
E_{\rm vac} = -\frac{1}{2}\sum_j D_j|\epsilon_j| - 
\frac{1}{2}\int_0^\infty dk \sqrt{k^2+m_f^2}\, \Delta\rho(k) + E_{\rm ct} \, .
\label{Evac2}
\end{equation} 
The momentum integral and $E_{\rm ct}$ are both divergent, but 
their sum is finite because the theory is renormalizable.  
We render the integral finite by subtracting the first $N$ 
terms in the Born series expansion of the density of states and adding back in 
exactly the same quantity in the form of Feynman diagrams:
\begin{eqnarray}
E_{\rm vac} & = & -\frac{1}{2}\sum_j D_j|\epsilon_j| 
- \frac{1}{2}\int_0^\infty dk \sqrt{k^2+m_f^2} 
\left( \Delta\rho(k) - \sum_{i=1}^{N}\Delta\rho^{(i)}(k)\right) 
\nonumber\\ & &  
+ \sum_{i=1}^N E_{\rm FD}^{(i)} + E_{\rm ct} \, .
\end{eqnarray}
When we combine $E_{\rm ct}$ with $\sum_{i=1}^N E_{\rm FD}^{(i)}$, 
we cancel the divergences as well as implement the renormalization prescription.
As a result, the above expression is manifestly finite.
The minimal number of required Born subtractions, $N$, 
is easily determined by an analysis of the superficial degree of 
divergence of the one-loop Feynman diagrams.  For our theory, $N=4$.

We will work with background fields in the spherical ansatz
\cite{RatraYaffe}.  Then we can express $\Delta\rho(k)$ (and its Born
series) in terms of the momentum derivative of the phase shifts \cite{Schwinger}, induced by the background fields, of the Dirac
wave-functions,
\begin{equation}
\Delta\rho(k) = \frac{1}{2\pi i}\frac{d}{dk}\Tr \ln S(k) 
= \frac{1}{\pi}\frac{d}{dk}\sum_{\sigma\in\{+,-\}}
\sum_G D_G\delta_{G,\sigma}(k) \, .
\end{equation}
Here we have expanded the $S$-matrix in grand spin channels labeled
by $G$.  The grand spin is the vector sum of total angular momentum and isospin. It is conserved by the potential, eq.~(\ref{potential}).
The asymptotic scattering states are labeled 
by parity $(-1)^G$ and total spin $G\pm1/2$ and we obtain a 
$4\times4\,S$-matrix in general (except in the $G=0$ channel, where it is $2\times2$). We let $\delta_{G,\sigma}(k)$ denote the 
sum of the eigenphase shifts 
at momentum $k$ in channel $G$ and $\sigma=\pm$ specify the sign of the energy 
eigenvalue, $\omega=\pm\sqrt{k^2+m_f^2}$.   Note that the single particle spectrum is not symmetric because the Dirac Hamiltonian is not charge conjugation invariant.  The degeneracy is given by $D_G=2G+1$.  In Appendix B we show in detail
how to use the Dirac equation to calculate the bound state energies and
scattering phase shifts needed in the computation of $E_{\rm vac}$.  

To simplify the calculation we only subtract the first two 
Born approximants and eliminate the remaining log-divergence in 
the momentum integral by using a limiting function approach 
developed in ref. \cite{DecouplingNoGauge}.  The final expression for the vacuum 
polarization energy is
\begin{equation}
E_{\rm vac} =  -\frac{1}{2}\sum_j(2G_j+1)|\epsilon_j| - 
\int_0^\infty \frac{dk}{2\pi}\,\sqrt{k^2+m_f^2}\,
\frac{d}{dk}\,\overline{\delta}(k) + E^{(1,2)} + E^{(3,4)} \, , 
\label{Evac}
\end{equation}
where $G_j$ is the grand spin associated
with the bound state $j$ and
\begin{equation}
\overline{\delta}(k) =  
\sum_{\sigma\in\{ +,- \}}\sum_{G=0}^\infty (2G+1)
\left(\delta_{G,\sigma}(k) - \delta^{(1)}_{G,\sigma}(k) - 
\delta^{(2)}_{G,\sigma}(k) \right) + \delta_{\rm lim}(k) \, .
\end{equation}
Here $\delta^{(i)}_{G,\sigma}(k)$ denotes the 
$i^{\rm th}$--term in the Born series of $\delta_{G,\sigma}(k)$.
After subtracting  $\delta^{(1)}$ and $\delta^{(2)}$ from
$\delta$, the momentum integral does not contain any
contributions that are linear or quadratic in $V$. 
The limiting function for the sum over all 
eigenphase shifts, $\delta_{\rm lim}(k)$,  eliminates 
the logarithmically divergent pieces
that are third and fourth order in $V$ from the momentum integral. Its 
analytic expression can be extracted from the divergent pieces
of the corresponding Feynman diagrams and
is given in Appendix B, eq.~(\ref{LimitingPhaseshift}).  Furthermore, $E^{(1,2)}$ denotes the 
contribution up to second order in $V$ from the renormalized Feynman
diagrams. Its explicit expression is displayed in Appendix B.
Finally, $E^{(3,4)}$ contains the third and fourth order
counterterm contribution combined with the divergences in the third
and fourth order Feynman diagrams that have been subtracted
from the momentum integral via $\delta_{\rm lim}$. Again, its 
explicit form can be found in Appendix B.

Thus we compute an exact, finite, renormalized, gauge-invariant 
effective energy functional, $E_{\rm eff}[A,\Phi]$, defined in eq.~(\ref{Eeff}), for the \HiggsGauge sector, with 
the fermion fields integrated out.   
\section{The Energy of a Fermionic Configuration}
We are interested in exploring the possibility of the emergence of a
stable, fermionic soliton in the \HiggsGauge sector of the theory as 
we increase the Yukawa coupling (thereby making the perturbative fermion heavier).  In the previous section we outlined
the procedure that allows us to calculate the effective energy when
the fermions are integrated out.  Now we analyze the minimum
additional energy required to associate unit fermion number with a particular
\HiggsGauge configuration $C$, where $C\equiv \{A,\Phi \}$.  
First in
section \ref{The Fermion Number of a Configuration} we determine
the integer fermion number $F[C]$ of a configuration.  (This is subtle because we have to contend with the anomalous violation of fermion number).    Then we
occupy or empty levels of the single-particle Dirac Hamiltonian
to obtain the lowest energy state with net fermion number 1.
If $F[C]=0$, then the lightest 
positive bound state needs
to be filled and the occupation energy $E_{\rm occ}^{(1)} =
\epsilon_{\rm lowest}$, where the superscript `(1)' denotes that levels have been occupied/emptied to obtain fermion number 1.  If $F[C]=1$ then $E_{\rm occ}^{(1)}=0$ because $C$
is already fermionic, and so on.  Thus, the minimum total energy of a
single fermion associated with a configuration is
\begin{equation}
E\oneF[C] = E_{\rm cl}[C] + E_{\rm vac}[C]+ E_{\rm occ}^{(1)}[C] \, .
\label{allTheEs}
\end{equation}

In previous works, such as \cite{NolteKunz}, the vacuum polarization
contribution was omitted from the above equation and stable
non-topological solitons were found.  We consider such calculations
inconsistent, because $E_{\rm occ}^{(1)}$ and $E_{\rm vac}$ are both order $\hbar$.
We will see explicitly in section \ref{Twisted Higgs} that $E_{\rm vac}$
makes a significant positive contribution when the Yukawa coupling is large enough that the perturbative fermion mass is comparable to the classical energy.

Since we refer to all these different energies frequently in the rest of the paper,  we summarize our notation in Table \ref{EnergiesTable}.\\
\begin{table}[hbt]
	\begin{tabular}{|c|l|} \hline
	\, $E_{\rm cl}$ \, & \, Classical Higgs and gauge energy, eq.~(\ref{Eclassical}) \\ \hline
	\, $E_{\rm vac}$ \, & \, Fermion vacuum polarization energy, including counterterms, eqs.~(\ref{Evac1}, \ref{Evac2}) \\ \hline
	\, $E_{\rm eff}$ \, & \, One--fermion--loop effective energy, $E_{\rm cl} + E_{\rm vac}$ \\ \hline
	\, $E_{\rm occ}^{(m)}$ \, & \, Smallest occupation energy required to obtain fermion number $m$, eq.~(\ref{allTheEs}) \\ \hline  
	\, $E_{\rm eff}^{(m)}$ \, & \, Smallest effective energy in the fermion number $m$ sector, $E_{\rm eff} + E_{\rm occ}^{(m)}$ \\ \hline
	\end{tabular}
\caption{\label{EnergiesTable} Definitions of some of the energies which appear in our analysis.}
\end{table}
\subsection{The Fermion Number of a Configuration}
\label{The Fermion Number of a Configuration}
First we review properties of the \HiggsGauge configuration space 
and the classical energy functional defined on it. From the 
expression for $E_{\rm cl}$ in eq.~(\ref{Eclassical}), it follows 
that configurations
\begin{equation}
A_j  =  \frac{i}{g}U^{(n)}\partial_j{U^{(n)}}^\dag \, , \\
\Phi  =  vU^{(n)} \, ,
\label{Vacua3D}
\end{equation}
have $E_{\rm cl}=0$ and we refer to them as {\it \classicalVacuum configurations}. Here $U^{(n)}$ is any map from $S^3$ to $SU(2)$ with winding 
number $n$.  \ClassicalVacuum configurations with the same winding number are equivalent 
under small (winding number 0) gauge transformations.
We use ${\mathcal C}^{(n)}$ to 
denote the homotopic class of \classicalVacuum configurations with 
winding number $n$.
Topologically inequivalent \classicalVacuum 
configurations are related by large (nonzero winding number)
gauge transformations. Along any continuous interpolation between two configurations, one in ${\mathcal C}^{(n)}$ and
 the other in ${\mathcal C}^{(m)}$ (with $n\ne m$), there exists a configuration, $C$, with maximum classical energy.  Since no $U^{(n)}$ can be
continuously deformed into any $U^{(m)}$, $E_{\rm cl}[C] > 0$.  The configuration corresponding to the minimax of these energies, when all interpolations are considered, is the {\em classical sphaleron}.  When the fermion vacuum polarization
energy is added to the classical energy to obtain the effective energy ($E_{\rm eff}=E_{\rm cl}+E_{\rm vac}$), not only does the magnitude of the minimax energy change,
but its location in configuration space shifts as well.  We therefore define the {\em quantum-corrected sphaleron} to
be the configuration that has the lowest of the maximum effective energies 
along all interpolations between topologically inequivalent \classicalVacuum
configurations. 

We associate any configuration $C$ with a unique class of
\classicalVacuum configurations by continuously deforming $C$ in the direction of the
negative gradient of the classical energy until we get a configuration in ${\mathcal C}^{(n)}$ for some $n = n(C)$.  We call ${\mathcal C}^{(n(C))}$ the 
{\it connected ${\mathcal C}$-class} of $C$ and we say that $C$ is 
in the winding number $n$ basin.  Note that the classical sphaleron and all configurations that descend to it are on the boundary between different basins.

For any two configurations $C_1$ and $C_2$, we define the 
{\it spectral flow} $S[C_1, C_2]$ to be the number of 
eigenvalues (levels) of the single particle Dirac 
Hamiltonian that cross zero from above 
minus the number that cross zero from below 
along any interpolation from $C_1$ to $C_2$.  The fermion number of a configuration $C$ is defined as
\begin{equation}
F[C] = S[C_1   , C]  
\label{fermionNumber}
\end{equation}
with $C_1 \in {\mathcal C}^{(n(C))}$.  Since the Dirac spectrum is gauge-invariant, $F[C]$ does not depend on which particular $C_1$ is chosen from the connected ${\mathcal C}$-class.  Moreover, $F[C]$ is gauge-invariant even under large gauge transformations.  Also, $F[C]$ does not depend on the chosen interpolation because the spectral flow is the same for all interpolations.  This definition of the fermion number can be readily understood for a C 
that has classical energy less than the energy 
of the classical sphaleron. A continuous interpolation from any configuration in ${\mathcal C}^{(n(C))}$ to $C$ preserves net fermion number 
because anomalous fermion number violations require the \HiggsGauge fields to 
cross the sphaleron barrier.  Thus, defining 
\classicalVacuum configurations to have zero
fermion number leads to eq.~(\ref{fermionNumber}).  Configurations
that have classical energies larger 
than the classical sphaleron are not separated by an
energy barrier from topologically inequivalent basins, so it is not
clear what their fermion number should be, although our  definition
does assign a unique $F$ to them.  

Having determined the fermion number of a configuration, we can use
eq.~(\ref{allTheEs}) to find $E\oneF $, the minimum effective energy of a configuration in the fermion number 1 sector.
\subsection{Stability of the Soliton}
We would like to know if there exists a configuration at
which the one-fermion effective energy functional, $E\oneF $, has a local
minimum.  This configuration would be a fermionic soliton.  We carry out a variational search, looking for a
configuration $C$ such that $E\oneF [C]<m_f$ and
$E\oneF [C]<E_{\rm q.s.}$, where $E_{\rm q.s.}$ is the effective energy of the quantum-corrected sphaleron.  The first condition ensures that
$C$ cannot simply decay into a configuration with
zero classical energy plus a perturbative fermion.  The second condition prevents $C$ from rolling over the  quantum-corrected sphaleron, giving up its
fermion number and then rolling down the $E\zeroF $ surface to a
\classicalVacuum configuration.  Finding a configuration with these
properties would guarantee the existence of a nontrivial
local minimum of $E\oneF $.
\section{The Search for the Soliton}
In this section we describe our search for the soliton.  
We first review the spherical ansatz for the gauge and Higgs fields.  
We then outline the restrictions imposed on the variational ans\"atze 
used to search for a soliton.  Finally, we report on our search within 
two physically motivated sets of ans\"atze: the ``twisted Higgs'' and ``paths 
over the sphaleron''.  Note that throughout this section the 
perturbative fermion mass is set to 1 so that energies 
and lengths are measured in units of 
$m_f$ and $1/m_f$, respectively. 
\subsection{The Spherical Ansatz}
We only consider static gauge and Higgs fields in the spherical ansatz.   This enables us to expand the fermion S matrix in terms of partial waves labeled by the grand spin $G$, as described in Appendix 
B.  Our method for calculating the fermion vacuum polarization energy requires
such an expansion. Under these restrictions (and in the $A_0 = 0$ gauge, which for smooth fields is obtained by a non--singular gauge transformation), the fields
can be expressed in terms of five real functions of $r$:
\begin{eqnarray}
A_i (\vec{x}) & = & -A^i(\vec{x})=\frac{1}{2g}\left[ a_1(r)\tau_j\hat{x}_j\hat{x}_i + 
\frac{\alpha(r)}{r}(\tau_i - \tau_j\hat{x}_j\hat{x}_i ) + 
\frac{\gamma(r)}{r}\epsilon_{ijk}\hat{x}_j\tau_k \right] \, , 
\nonumber \\
\Phi (\vec{x}) & = & v \left[ s(r) + ip(r)\tau_j\hat{x}_j \right] \, ,
\label{SphericalAnsatz}
\end{eqnarray}
where $\hat{x}$ is the unit three-vector
in the radial direction.  

The ansatz transforms under a $U(1)$ subgroup of
the full $SU(2)$ gauge symmetry, corresponding to elements of the 
form
\begin{equation}
g(\vec{x}) = e^{if(r)\tau_j\hat{x}_j/2} \, ,
\label{sphGauge}
\end{equation}
with $a_1$ transforming as a 1 dimensional vector field, $s+ip$
as a scalar with charge $1/2$, and $\alpha + i(\gamma-1)$ as a
scalar with charge $1$.  It is thus convenient to introduce the moduli $\rho, \Sigma$ and phases $\theta, \eta$ for the charged scalars:
\begin{equation}
-i\rho e^{i\theta} \equiv \alpha + i(\gamma-1)
\quad{\rm and}\quad
\Sigma e^{i\eta} \equiv s + ip   \, .
\label{polarfields}
\end{equation}
From now on we will specify a configuration using the five functions $a_1(r), \rho(r), \theta(r), \Sigma(r)$ and $\eta(r)$.  Regularity of $A_i(\vec{x})$ and $\Phi(\vec{x})$ at $\vec{x}=0$ requires that
\begin{eqnarray}
\rho(0) & = & 1 \, , \nonumber \\
\rho'(0) & = & 0 \, , \nonumber \\
\theta(0) & = & -2n_{\theta}\pi \, , \nonumber \\ 
a_1(0) & = & \theta '(0) \, ,  \nonumber \\
\mbox{either } \Sigma(0) & = & 0 \mbox{ or } \eta(0) = -n_{\eta}\pi \, .
\label{BoundCond0}
\end{eqnarray} 
Here $n_{\theta}, n_{\eta}$ are integers and primes denote derivatives with respect to the radial coordinate.

For the gauge transformation $g(\vec{x})$ in eq.~(\ref{sphGauge}) to be non-singular, we require $f(0)=-2n\pi$, where $n$ is an integer and we denote this boundary condition as a superscript: $f(r) \equiv f^{(n)}(r)$.  If $f^{(n)}(r)$ is restricted to be $0$ as $r\rightarrow\infty$ (which is equivalent to
$g(r\rightarrow \infty)=\ID$) then $n$ is the winding of the map
$g(\vec{x}):S^3\rightarrow SU(2)$.  So the topology of the \classicalVacuum configurations persists in the spherical ansatz. The classical energy of eq.~(\ref{Eclassical}) takes the form 
\begin{eqnarray}
E_{\rm cl} & = & 4\pi \int_0^\infty dr 
\left\{ \frac{1}{g^2}\left[\rho^{\prime^2}
+ \rho^2(\theta^\prime  - a_1)^2 
+ \frac{(\rho^2-1)^2}{2r^2}  \right]  \right. 
\nonumber \\ &&\hspace{2cm} 
\left. +\frac{1}{f^2}\left[ r^2 \Sigma^{\prime 2 }
+ r^2\Sigma^2(\eta^\prime  - \frac{1}{2}a_1)^2 
+\frac{r^2}{4}m_h^2(\Sigma^2-1)^2  
 \right. \right.  \nonumber \\ && \hspace{3cm} 
\left. \left. + \frac{1}{2}\Sigma^2\left( (\rho - 1)^2 
+ 4\rho^2\sin^2\frac{\theta-2\eta}{2} \right) \right]\right\} \, ,
\label{eclchiral}
\end{eqnarray} 
and winding number $n$ \classicalVacuum configurations of eq.~(\ref{Vacua3D}) now become 
\begin{eqnarray}
\rho(r)&=&1\, , \quad \Sigma(r)=1 \, , \nonumber \\
\theta(r)&=&f^{(n)}(r)\, , \quad
\eta(r)=\frac{f^{(n)}(r)}{2}\, , \quad
a_1(r)=f^{(n)\prime}(r)\, .
\label{Vacua1D}
\end{eqnarray}

We want the \HiggsGauge fields to have finite classical energy.   So we require a field configuration of the form eq.~(\ref{Vacua1D}) as $r\rightarrow\infty$, and the restriction that $f^{(n)}(\infty)=0$ uniquely specifies the boundary conditions on the fields at infinity.  At $\vec{x}=0$, the boundary conditions on $\rho$ (specified in eq.~(\ref{BoundCond0})) make the energy density finite and we do not require any additional constraints. 

The anomalous violation of fermion number is given by the anomaly equation
\begin{equation}
\partial_\mu \left(\overline{\Psi}\gamma^\mu\Psi\right) 
= \partial_\mu K^\mu \, ,
\end{equation}   
where $K^\mu$ is the Chern-Simons current.  It is useful to consider the charge associated with it:
\begin{eqnarray}
N_{\rm CS} & = & -\frac{g^2}{8\pi^2}\epsilon_{ijk}\int d^3x 
\tr \left(A_i\partial_jA_k-\frac{2}{3}igA_iA_jA_k  \right) 
\nonumber \\
& = & \frac{1}{2\pi}\int_0^\infty dr 
\left[a_1+\rho^2(\theta^\prime - a_1)\right] \, .
\end{eqnarray}
This is a non-integer in general, and is equal to the integer winding
number of $f^{(n)}$ for the 
configurations of eq.~(\ref{Vacua1D}), and a half-integer for the
sphaleron \cite{KlinkhamerManton}.  Under a winding $n$ gauge transformation, $N_{\rm CS}
\rightarrow N_{\rm CS}+n$.  For background fields that interpolate between
topologically distinct \classicalVacuum configurations, the net
fermion number produced is given by the change in $N_{\rm CS}$.
\subsection{Restrictions on the Variational Ans\"atze}
\label{Restrictions}   
Our methods allow us to consider any static, spherically symmetric
configuration, $C$, in the \HiggsGauge sector, specified by 
the five real
functions $a_1(r)$, $\rho(r)$, $\theta(r)$, $\Sigma(r)$ and
$\eta(r)$.  In principle, we could numerically minimize the fermionic
energy, $E\oneF [C]$, in terms of the five functions and determine if
a soliton exists.  In practice however, that is not feasible.
So instead we limit ourselves to the variation of a few parameters in
ans\"atze motivated by physical considerations.

We restrict our variational ans\"atze to those that obey the 
above described boundary conditions at the origin 
and at infinity.
In addition, we restrict the
Higgs fields to lie within the chiral circle, $\Sigma(r)<1$, because
otherwise the effective potential is
unbounded from below.  (The leading terms in the derivative expansion of eq (10) can be found in \cite{Ball:xg}).  Finally, the effective theory (obtained by integrating out the fermions) has a Landau pole in the ultraviolet, reflecting new dynamics at some
cutoff energy scale or equivalently at a minimum distance scale.
Configurations that are large compared to this distance scale 
are relatively
insensitive to the new dynamics at the cutoff, but smaller
configurations are sensitive.  For small widths and large couplings,
the Landau pole becomes significant and leads to unphysical negative
effective energies in eq.~(\ref{Eeff}) \cite{Ripka:ne}.  We have to be wary of this in
estimating the reliability of our results.  See
\cite{DecouplingNoGauge} for more detailed discussions on the
effective potential and the Landau pole.
\subsection{Twisted Higgs}
\label{Twisted Higgs}
We first consider twisted Higgs configurations, with $n_{\eta}=1$ so that
$\eta(r)$ goes from $-\pi$ at $r=0$ to $0$ as
$r\rightarrow\infty$.  The other functions are trivial: $a_1(r)=0, \rho(r)=1, \theta(r)=0$ and $\Sigma(r)=1$.  If we smoothly interpolate from a
\classicalVacuum configuration (in the connected ${\mathcal C}$-class)
to such a configuration, we observe that one fermion bound state, that originates in the positive continuum, has its energy decrease sharply.  The wider the final twisted
Higgs configuration, the closer this level ends up to the negative
continuum.  At a width of order $1/m_f$, it has energy zero,
eliminating  any occupation energy contribution, $E_{\rm occ}^{(1)}$, to the
fermionic energy, $E\oneF $, associated with it.  The existence of
this level makes such twisted Higgs configurations attractive
candidates for the variational search.

We consider one such twisted Higgs configuration with a width
characterized by a variational parameter $w$,
\begin{equation}
\eta =  -\pi e^{-r/w}
\label{twistedHiggs}
\end{equation}
and add various perturbations to it.  For instance, one among many of our variational ans\"atze (in the $\theta=0$ gauge) is a four parameter ansatz with parameters $p_0,\ldots,p_3$: 
\begin{eqnarray}
\eta & = & -\pi e^{-r/w} + p_0 \frac{r/w}{1+(r/w)^2}e^{-r/w} \, , \nonumber \\
\Sigma & = & 1 + p_1 \frac{1}{1+(r/w)}e^{-r/w} \, , \nonumber \\
a_1 & = & p_2 \frac{r/w}{1+(r/w)^2}e^{-r/w} \, , \nonumber \\
\rho & = & 1 + p_3 \frac{(r/w)^2}{1+(r/w)^3}e^{-r/w} \, , 
\end{eqnarray}
where $-1<p_1<0$ (to keep the Higgs field within the chiral circle and
its magnitude positive) and $p_3>-5.23$ (to keep $\rho$ positive).  
For a prescribed set of theory parameters ($m_w, m_h$ and $f$) we determine the gauge coupling $g=\sqrt{2}m_w^{(0)}/v$ from the renormalization constraint eq.~(\ref{ModelParamsConstraint}).  We then vary the ansatz parameters ($w, p_0, \ldots , p_3$) 
to lower the fermionic energy $E\oneF $.  We find
that the gain in binding energy is insufficient to compensate for the
increase in the effective energy $E_{\rm eff}$.   Through all our
variations, $E\oneF $ is strictly greater
than $m_f$ and we find no evidence for a soliton.  The same result was
obtained in \cite{DecouplingNoGauge} without gauge fields, and the
extra gauge degrees of freedom do not seem to help in the twisted
Higgs ansatz.

\begin{figure}[htb]
\includegraphics{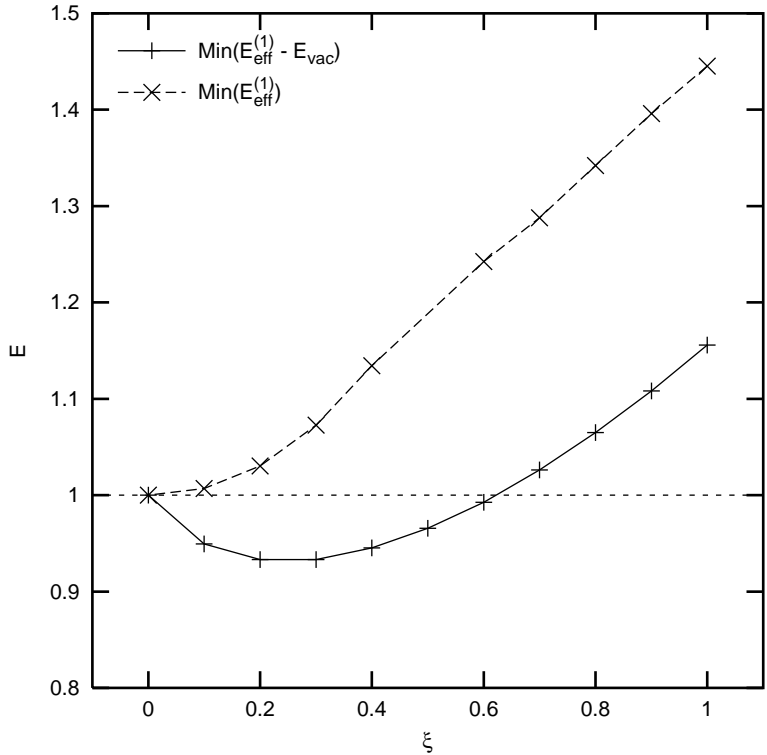}
\caption{\small Minimum fermionic effective energies (in units of $m_f$), with as well as without $E_{\rm vac}$ contributions, along the interpolation in eq.~(\ref{twistedHiggsInterpolation}). }
\label{soliton}
\end{figure}

We find that the fermion vacuum polarization contribution, $E_{\rm vac}$, to the fermionic energy, $E\oneF $ destabilizes would-be solitons.  Consider a linear interpolation from the trivial \classicalVacuum configuration to the twisted Higgs configuration in eq.~(\ref{twistedHiggs}) with gauge fields set to zero.  We introduce the interpolating parameter $\xi$ which goes from 0 to 1:
\begin{equation}
\Sigma e^{i\eta} = 1 - \xi + \xi \exp\left( -i\pi e^{-r/w} \right) \, .
\label{twistedHiggsInterpolation}
\end{equation}  
We choose the Yukawa coupling to be $f=10$ and the Higgs mass to be $v/\sqrt{2}$.  Since the gauge fields are trivial, the classical energy as well as the Dirac spectrum are independent of the gauge coupling $g$ and the gauge bosons mass $m_w$.  For each value of $\xi$, we compute $E\oneF $ and $E\oneF  - E_{\rm vac}$ for different values of the width parameter $w$.  In Fig. \ref{soliton} we plot $E\oneF $ and $E\oneF  - E_{\rm vac}$ as functions of $\xi$, choosing the width at every point to minimize the energy (we do not allow $w$ to be less than 1 so that we remain relatively insensitive to the Landau pole).  For all points on the plot there is no spectral flow and so $E\oneF  = E_{\rm cl} + \epsilon_{\rm lowest} + E_{\rm vac}$ in accordance with eq.~(\ref{allTheEs}), where $\epsilon_{\rm lowest}$ is the smallest positive bound-state energy in the Dirac spectrum.  If $E_{\rm vac}$ is omitted, for $0<p<0.6$ we have configurations that have fermionic energies lower than the mass of the perturbative fermion ($m_f = 1$ in our units).  These configurations indicate the existence of a local minimum on the $E\oneF -E_{\rm vac}$ surface which would be a soliton.  The $E_{\rm vac}$ contribution, however, raises the energies of the configurations to above $m_f$ as shown in the figure, and the would-be solitons are destabilized.
\subsection{Paths over the Sphaleron}
\label{Paths over the Sphaleron}
The gauge fields introduce another mechanism for strongly binding a
fermion state: there is a zero mode in the background of the
sphaleron.  Along an interpolation of the background fields from a
configuration in ${\mathcal C}^{(n)}$ to a configuration in ${\mathcal C}^{(n+1)}$, a
fermion level leaves the positive continuum, crosses 
zero from above and
finally enters the negative continuum.  The lowering of the occupation
energy, $E_{\rm occ}^{(1)}$, as we approach zero from above is offset by
the rising effective energy $(E_{\rm eff}=E_{\rm cl}+E_{\rm vac})$, so
we must investigate whether the former can dominate the latter.  We also use
such interpolations to study the effects of a large Yukawa coupling on the sphaleron.  We approximate the
quantum-corrected sphaleron by minimizing the effective energy
barrier between topologically inequivalent \classicalVacuum
configurations, with respect to 
the variational parameters of our interpolations.  We also 
investigate the possible emergence of new barriers in the 
one-fermion energy surface when the perturbative fermion becomes heavier than the 
quantum-corrected sphaleron.  
These last two phenomena affect the stability of the heavy 
fermion and in some models may be significant for baryogenesis.
  
We make the following choices for the theory parameters: we fix the
Yukawa coupling at $f=10$, which is large enough that fermion effects
are significant, but small enough to prevent our configurations
from being affected by the Landau pole.  Indeed we encounter no
negative energy instabilities in our computations of 
$E_{\rm vac}$ for this coupling.  
We keep the Higgs mass fixed at $v/\sqrt{2}$,
which corresponds to $m_h = \frac{m_f}{\sqrt{2}f} \approx 0.07 m_f$.
We choose $g$ to keep the quantum-corrected sphaleron energy
comparable to $m_f$, since it is plausible that the mass of the
sphaleron sets the scale for decoupling.  If the sphaleron is
too heavy compared to the fermion, the binding energy gained by
the level crossing would be washed out by the effective energy of the
sphaleron.  For $g=6.5$ our best approximation to the
quantum-corrected sphaleron is approximately degenerate with the
fermion. When the gauge coupling is given, the mass $m_w$ of the gauge boson is determined from the
renormalization constraint eq.~(\ref{ModelParamsConstraint}):
$m_w \approx 0.63 m_f$ for $g=6.5$ and $m_w\approx0.98 m_f$ for $g=10$.
These theory parameters are of course large deviations from the
Standard Model parameters. We exaggerate them to 
see the effects of the heavy perturbative fermion.  Another concern is
that for large $g$, we should consider quantum fluctuations of the \HiggsGauge fields. However, we believe that anomaly cancellation drives the creation of a soliton, which would suggest that the fermion vacuum polarization
contains the essential physics, and our methods allow us to exactly
compute this contribution to the energy for any Yukawa coupling.

First we consider a linear interpolation between a 
winding-0 and a winding-1 \classicalVacuum 
configuration. The interpolation parameter $\xi$ goes from 0 to 1:
\begin{eqnarray}
\Phi & = & v(1-\xi)\ID + \xi v U^{(1)} \, , \nonumber \\
A_j & = & \xi \frac{i}{g}U^{(1)}\partial_j{U^{(1)}}^{\dag} \, .
\label{linearpath1}
\end{eqnarray}
In the spherical ansatz, $U^{(1)}(\vec{x})=g(\vec{x})$ is specified by a single 
function, as in eq.~(\ref{sphGauge}), that we choose to be 
\begin{equation}
f^{(1)}(r) = -2\pi e^{-r/w} \, ,
\label{linearpath2}
\end{equation}
where $w$ characterizes the width of the configuration.  
We interpolate $\xi$ from 0 (the trivial configuration with $N_{\rm CS}=0$) to 1/2 (a configuration with
$N_{\rm CS}=1/2$ in the presence of which the fermion has a zero mode) and
vary $w$ along the interpolation to minimize $E\oneF $.  This gives
an upper bound on the minimum $E\oneF $ as a function of $N_{\rm CS}$.
For $N_{\rm CS}=1/2$, this is an upper bound on the quantum-corrected
sphaleron energy as well, because $E\oneF  = E_{\rm eff}$ in the presence of
a fermion zero mode (the occupation energy is then 0)
\footnote{Our numerical methods do not allow us to consider 
$N_{\rm CS}$ exactly equal to $1/2$, because the Higgs
magnitude vanishes at $r=0$ and the second order Dirac equations
develop a singularity (as discussed in Appendix B), 
but we can
compute $E\oneF $ very close to $N_{\rm CS}=1/2$.}.  An exploration of
$N_{\rm CS}$ between 0 and 1/2 is sufficient to map to all values of
$N_{\rm CS}$, since configurations with $N_{\rm CS}$ between 1/2 and 1 are 
obtained by
charge conjugation, and configurations with $N_{\rm CS}<0$ or $N_{\rm CS}>1$ are
large-gauge-equivalent to configurations with $N_{\rm CS}$ between 0 and 1.

We also consider an instanton-like configuration where the Euclidean
time $\xi=x_4$ is the interpolation 
parameter (which varies from $-\infty$ to $\infty$) between two topologically
inequivalent \classicalVacuum configurations:
\begin{eqnarray}
A_\mu & = & h(r,\xi)\frac{i}{g} 
U_{\rm inst}(\vec{x},\xi)\partial_\mu 
U^{\dag}_{\rm inst}(\vec{x},\xi) \, , \nonumber \\
\Phi & = & v \sqrt{h(r,\xi)}\,U_{\rm inst}(\vec{x},\xi) \, ,
\label{definst1}
\end{eqnarray}
where 
\begin{equation}
U_{\rm inst}(\vec{x},\xi)=
\frac{\xi + i \tau_{j}x_{j}}{\sqrt{r^2+\xi^2}}
\label{definst2}
\end{equation}
is the canonical winding-1 map from $S^3$ (space-time infinity) to
$SU(2)$.  Furthermore $h$ is a function of the 
Euclidean space-time radius ($\sqrt{r^2+\xi^2}$)
and goes from 0 to 1 as this radius goes from 0 to $\infty$.
't Hooft's electroweak instanton \cite{'tHooft} is
constructed as a self--dual gauge field configuration
in the topological charge one sector,
and a Higgs field configuration that minimize the covariant kinetic
term in the Lagrangian density.  This gives
\begin{equation}
h(r,\xi)=\frac{r^2+\xi^2}{r^2+\xi^2+w^2}
\end{equation}
for any width $w$ (the classical theory with no Higgs field is
scale-invariant).  
We modify this radial function to exponentially
approach its asymptotic value of 1, so that the potential in our Dirac
equation falls off fast enough to have a 
well-defined scattering problem (as described in Appendix B).   We choose
\begin{equation}
h(r,\xi) = 1-e^{-(r^2+\xi^2)/w^2} \, .
\label{definst3}
\end{equation}
This choice does not minimize any part of the classical Euclidean
action (in the topological charge one sector). Since we are interested  in minimizing $E\oneF $,
which has fermion vacuum energy and occupation energy contributions in
addition to the \HiggsGauge sector classical energy, we do not need our configurations to minimize the classical energy.  In fact, as we
describe later, the configurations that minimize $E\oneF $ are
rather different from those that minimize $E_{\rm cl}$.

In order to compute the Dirac spectrum for this background 
using the methods described in Appendix B, 
we gauge transform 
to $A_0=0$ and 
$\lim_{r\rightarrow\infty}\eta=\lim_{r\rightarrow\infty}\theta=0$ 
using the transformation function
\begin{equation}
f(r,\xi) = \int_{-\infty}^{\xi}d\xi^\prime \frac{2r}{r^2+(\xi^\prime)^2}
h(r,\xi^\prime) - 2\pi
\end{equation}
in eq.~(\ref{sphGauge}).
Finally, as in the case of the linear interpolation, 
we consider $\xi$ from $-\infty$ (the trivial configuration with 
$N_{\rm CS}=0$) to 0 (a configuration with $N_{\rm CS}=1/2$ in the 
presence of which the fermion has a zero mode) and for each $\xi$ 
we choose the $w$ that minimizes $E\oneF $.

\begin{figure}[t]
\includegraphics{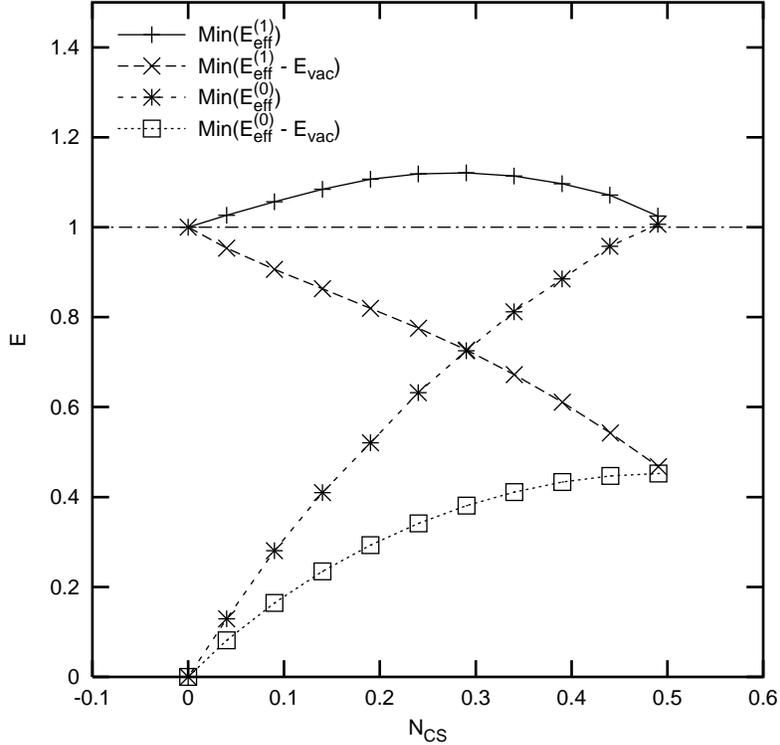}
\caption{\small Minimum effective energies (in units of $m_f$) along the linear path in eqs.~(\ref{linearpath1}, \ref{linearpath2}), in both the zero-fermion and one-fermion sectors (with as well as without the $E_{\rm vac}$ contributions).}
\label{quantum}
\end{figure}
In Fig.~\ref{quantum} we plot the minimum effective energies in both
the zero-fermion sector ($E\zeroF $) 
and the one-fermion sector
($E\oneF $), minimized within our variational ansatz for the linear
interpolation, as
functions of the Chern-Simons number, $N_{\rm CS}$ (see eqs.~(\ref{linearpath1},\ref{linearpath2})).  
As mentioned before, we fix the theory
parameters at a Yukawa coupling of $f=10$, a gauge coupling of $g=6.5$
and a Higgs mass of $m_h\approx 0.07 m_f$.
These parameters determine the mass of the gauge bosons to be $m_w
\approx 0.63 m_f$ from the renormalization constraint,
eq.~(\ref{ModelParamsConstraint}).  To isolate and highlight the
contribution of the fermion vacuum polarization energy, $E_{\rm vac}$,
we also plot the energies minimized with $E_{\rm vac}$ subtracted.

First consider the zero-fermion sector with the two curves $E\zeroF $
and $E\zeroF  - E_{\rm vac}$.  For all points on the plot there is
no spectral flow and so $E\zeroF  - E_{\rm vac}= E_{\rm cl}$.  At
$N_{\rm CS}=0$, both $E\zeroF $ and $E_{\rm cl}$ are minimized
at the trivial \classicalVacuum configuration, eq.~(\ref{Vacua1D}) with $f^{(0)}(r)=0$.  At $N_{\rm CS}=1/2$, $E\zeroF $
is minimized at the quantum-corrected sphaleron while $E_{\rm cl}$
is minimized at the classical sphaleron.  Within our variational
ansatz, we find the parameters that minimize $E\zeroF $ at
$N_{\rm CS}=1/2$ are different from those that minimize $E_{\rm cl}$.  So
our approximation to the quantum-corrected sphaleron is distinct from
our approximation to the classical sphaleron.  Moreover, the fermion
vacuum polarization energy correction to the sphaleron turns out to be
rather large.  
Our classical sphaleron has an energy of $0.45 m_f$ (which
agrees well with the numerical estimate of $E=1.52\frac{4\pi
v\sqrt{2}}{g}$ in \cite{KlinkhamerManton}), while our
quantum-corrected sphaleron has an energy of $1.02 m_f$. 

Next consider the $E\oneF $ and $E\oneF  - E_{\rm vac}$ plots in
the one-fermion sector in Fig.~\ref{quantum}.  
Again, for all points on the plot there is no
spectral flow and so $E\oneF  = E_{\rm cl}+ \epsilon_{\rm lowest} + E_{\rm vac}$
in accordance with eq.~(\ref{allTheEs}), where $\epsilon_{\rm lowest}$ is the
smallest positive bound-state energy in the Dirac spectrum.  Since the
classical sphaleron has an energy much smaller than the 
perturbative fermion mass, one would 
expect that the perturbative
fermion would have an unsuppressed decay mode over the
sphaleron, as first pointed out by Rubakov in \cite{Rubakov}. 
The
$E\oneF -E_{\rm vac}$ curve indeed displays this decay path.  The
fermion vacuum polarization energy modifies things in two crucial
ways.  First, the fermion quantum corrections to the sphaleron raise
its energy to be degenerate with the fermion, as mentioned before.  So
the threshold mass is significantly increased.  Second, in the plot of
$E\oneF $ we observe that there is an energy barrier 
between the fundamental fermion
and the quantum-corrected sphaleron.  This indicates that even when
the fermion becomes heavier than the sphaleron, there might exist a
range of masses for which the decay continues to be exponentially
suppressed (since it can proceed only via tunneling).

\begin{figure}[htb]
\includegraphics{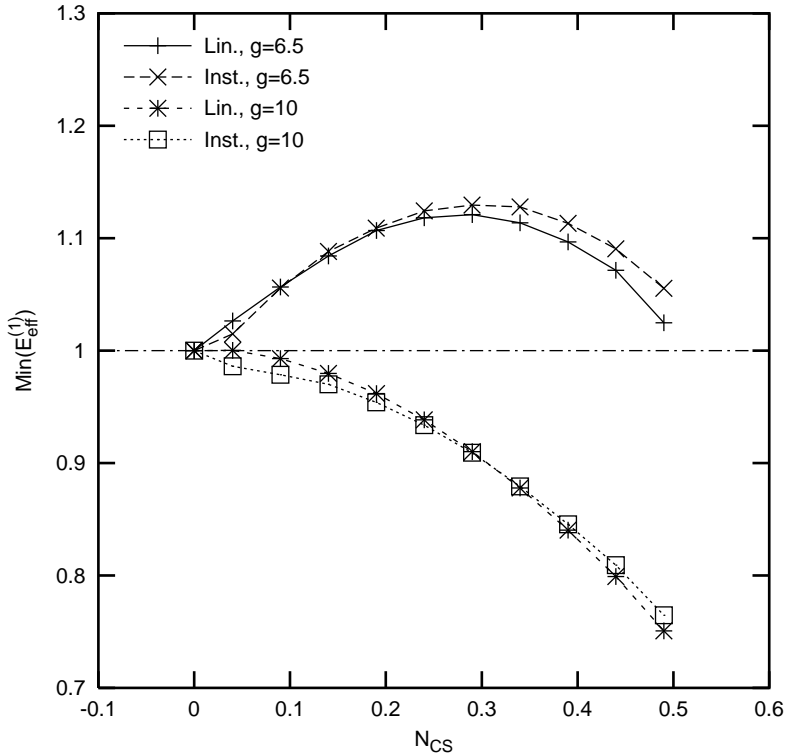}
\caption{\small Minimum $E\oneF $ (in units of $m_f$) for paths from 
a \classicalVacuum configuration to the sphaleron. Curves 
denoted 'Lin.' refer to the linear path in eqs.~(\ref{linearpath1}, \ref{linearpath2})
while those labeled 'Inst.' are associated with the instanton 
path, eqs.~(\ref{definst1},\ref{definst2},\ref{definst3}).}
\label{pathOverSph}
\end{figure}
In Fig.~\ref{pathOverSph} we restrict our attention to the 
one-fermion sector and consider the linear interpolation and the
instanton-like interpolation.  In addition to $g=6.5$ we consider $g=10$, which
corresponds to $m_w \approx 0.98 m_f$.  For each of these two gauge
couplings, we plot the effective energy minimized within our ans\"atze
as a function of $N_{\rm CS}$ for the two interpolations.

The two seemingly different interpolating configurations produce very
similar minimum $E\oneF $ curves.  Furthermore,
when we enlarge the variational ans\"atze in our interpolations, we
are unable to reduce the energies by any significant amount.  Thus we
speculate that the plotted curves may be close to tight upper bounds
on the true minimum $E\oneF $ curve.  This is the justification for
considering only the linear interpolation in Fig.~\ref{quantum} and
taking the evidence for the emergence of a new barrier and the
significant energy change of the sphaleron seriously.

Note that as the gauge coupling increases, lowering the energy
of the quantum-corrected sphaleron, the barrier between the
fundamental fermion and the sphaleron does not persist indefinitely.
We observe that for $g=10$, when $m_f$ is approximately 1.3 times our
quantum-corrected sphaleron, there is no barrier and the decay mode
is finally unsuppressed.
 
Just as in the case of the twisted-Higgs variational ansatz, in our
ans\"atze of paths from a \classicalVacuum configuration 
to the sphaleron,
we have not found a configuration with the associated fermion energy
lower than both the perturbative fermion and the quantum-corrected
sphaleron.  Thus, we find no evidence for the existence of fermionic
solitons in the low energy spectrum of the Standard Model within our
 ans\"atze.
\section{Conclusions and Discussion}
We have explored quantum effects of a heavy fermion on a chiral gauge theory.  The quantum-corrected sphaleron is heavier than the classical sphaleron by an energy of the order of the perturbative fermion mass.  This higher barrier suppresses fermion number violating processes.  We also observe the creation of an energy barrier between the perturbative fermion and the sphaleron, so that a fermion with energy
slightly above the sphaleron can still only decay through tunneling.  This new barrier exists only for an intermediate range of perturbative fermion masses, and a heavy enough perturbative fermion is indeed unstable.  We do not, however, find evidence for the existence of a soliton in the spectrum of the theory.  The fermion vacuum polarization contribution seems to destabilize any would-be solitons.  

It is possible that for a large enough Yukawa coupling, the Witten anomaly is saturated by states that do not have a particle interpretation.  But we believe the anomaly puzzle could still be resolved by soliton states that exist outside the spherical ansatz.  Although the reduced $U(1)$ theory of the spherical ansatz reproduces the anomalous violation of fermion number, it does not have the Witten anomaly.  This suggests that the ansatz might be too restrictive to resolve decoupling issues associated with the Witten anomaly.  Klinkhamer has conjectured the existence of
a non-spherically-symmetric sphaleron \cite{Klinkhamer} derived using the same topological
properties underlying the Witten anomaly, which would provide one candidate
background.  Like the usual sphaleron, this configuration has a zero
mode because it represents the middle of a path in which a fermion
level crosses zero.  However, an evaluation of the stability of such a state would require a
calculation of the one--loop effective energy, which is more difficult in the
absence of spherical symmetry.
\section*{ACKNOWLEDGMENTS}
We would like to thank J.~Goldstone, M.~Quandt, and  
F.~Wilczek for valuable discussions and also P.~Sundberg for his help. 
E.~F., R.~L.~J. and V.~K. are supported in part by the U.S.~Department
of Energy (D.O.E.) under cooperative research agreement~\#DF-FC02-94ER40818.
N.~G. is supported by the U.S.~Department of Energy (D.O.E.) under
cooperative research agreement~\#DE-FG03-91ER40662 and H.W. is
supported by the Deutsche Forschungsgemeinschaft under contract We
1254/3-2.

\appendix
\section{Results from Feynman Diagrams}

In this Appendix we list the results from the Feynman diagram
calculations mentioned in section III.

In dimensional regularization ($d=4-\epsilon$)
the counterterm coefficients, defined in eq.~(\ref{Lct}), read
\begin{eqnarray}
c_1 & = & \frac{1}{6}\frac{g^2}{(4\pi)^2} \Biggl[ {\mathcal D} - \frac{1}{2} 
- 3\int_0^1dx x(1-x) \Biggl( 2\ln\frac{\Delta(x,m_w^2)}{m_f^2} 
- x(1-x)\frac{m_w^2}{\Delta(x,m_w^2)}  
\Biggr)  \Biggr] \, , \nonumber \\
c_2 & = & -\frac{f^2}{(4\pi)^2}\left[ {\mathcal D} - \frac{2}{3} - 
6\int_0^1 dx x(1-x)\ln \frac{\Delta(x,m_h^2)}{m_f^2} \right] \, , 
\nonumber \\
c_3 & = & 2m_f^2\frac{f^2}{(4\pi)^2}\left({\mathcal D}+1\right) \, , 
\nonumber \\
c_4 & = & \frac{f^4}{4(4\pi)^2}\left[ 4{\mathcal D} - \frac{m_h^2}{m_f^2} 
- 6\int_0^1dx\ln\frac{\Delta(x,m_h^2)}{m_f^2} \right] \, .
\end{eqnarray}
We have introduced the abbreviations
\begin{equation}
\Delta(x,q^2) \equiv m_f^2 - x(1-x)q^2 \quad {\rm and}\quad
{\mathcal D} \equiv \frac{2}{\epsilon} - \gamma + \ln 
\frac{4\pi\mu^2}{m_f^2} \, ,
\end{equation}
where $\mu$ is the momentum scale introduced to
maintain the canonical dimensions of the parameters 
when regularizing in fractional dimensions. 

In eq.~(\ref{Evac}) $E^{(1,2)}$ denotes the contribution
to the vacuum polarization energy from first and second order 
renormalized Feynman diagrams. Its explicit expression reads
\begin{eqnarray}
E^{(1,2)} & = & 
\frac{-2}{(4\pi)^2}\int\frac{d^3q}{(2\pi)^3} 
\Biggl\{f^2\fourier{h}(\vec{q})\fourier{h}(-\vec{q}) 
\Biggl[ -(q^2+m_h^2) + 6\int_0^1 dx \Delta(x,-q^2)
\ln\frac{\Delta(x,-q^2)}{\Delta(x,m_h^2)} \Biggr]
\nonumber \\ && \hspace{2cm}
- m_f^2 \fourier{p}^a(\vec{q})\fourier{p}^a(-\vec{q}) 
\Biggl[q^2-6q^2\int_0^1 dx x(1-x) 
\ln\frac{\Delta(x,-q^2)}{\Delta(x,m_h^2)} 
\nonumber \\ && \hspace{7cm}
-2m_f^2\int_0^1 dx \ln\frac{\Delta(x,-q^2)}{m_f^2} \Biggr]  
\nonumber \\ & & \hspace{1cm}
+\frac{g^2}{2}\tr\left(\vec{q}\cdot\vec{\fourier{A}}(\vec{q})\vec{q}
\cdot\vec{\fourier{A}}(-\vec{q})  \right) 
\Biggl[\frac{1}{6} - 
2\int_0^1 dx x(1-x)\ln\frac{\Delta(x,-q^2)}{\Delta(x,m_w^2)}  
\nonumber \\ &&\hspace{7cm}
-\int_0^1 dx x^2(1-x)^2\frac{m_w^2}{\Delta(x,m_w^2)} \Biggr] 
\nonumber \\ & & \hspace{1cm}
+\frac{g^2}{2}\tr\left(\vec{\fourier{A}}(\vec{q})
\cdot\vec{\fourier{A}}(-\vec{q})\right)
\Biggl[-\frac{q^2}{6} - \frac{2}{3}m_f^2 
+ 2q^2\int_0^1 dx x(1-x)\ln\frac{\Delta(x,-q^2)}{\Delta(x,m_w^2)}  
\nonumber \\ & & \hspace{4cm}
+q^2\int_0^1 dx x^2(1-x)^2\frac{m_w^2}{\Delta(x,m_w^2)} 
+ m_f^2\int_0^1 dx \ln\frac{\Delta(x,-q^2)}{\Delta(x,m_h^2)}  
\nonumber \\ & & \hspace{4cm}
-5m_f^2\int_0^1 dx x(1-x)\ln\frac{\Delta(x,m_h^2)}{m_f^2} \Biggr] 
\nonumber \\ & & \hspace{1cm}
-igm_f^2\vec{q}\cdot\vec{\fourier{A}}^a(\vec{q})\fourier{p}^a(-\vec{q}) 
\Biggl[-\frac{2}{3}
+\int_0^1 dx \ln\frac{\Delta(x,-q^2)}{m_f^2}  
\nonumber \\ & & \hspace{5cm}
-6\int_0^1 dx x(1-x)\ln\frac{\Delta(x,m_h^2)}{m_f^2} \Biggr] \Biggr\} \, ,
\end{eqnarray}
with the Fourier transform of a field $\varphi(\vec{x})$ defined in the
usual way as $\fourier{\varphi}(\vec{q}) = \int d^3x
\varphi(\vec{x})e^{i\vec{q}\cdot\vec{x}}$.  The third and fourth order
counterterm contribution combined with the divergences in the third
and fourth order Feynman diagrams is
\begin{eqnarray}
E^{(3,4)} & = & \int \frac{d^3 x}{(4\pi)^2}\, \tr\, \Biggl\{ 
\frac{g^3}{6}\left(4i\partial_iA_j + g[A_i,A_j]\right)[A_j,A_i] 
\Biggl[\frac{1}{2} + 6\int_0^1 dx x(1-x)\ln\frac{\Delta(x,m_w^2)}{m_f^2}  
\nonumber \\ &&\hspace{7cm}
-3\int_0^1 dx x^2(1-x)^2\frac{m_w^2}{\Delta(x,m_w^2)}\Biggr] 
\nonumber\\ & & \hspace{1cm}
-g\vec{A}\cdot\left[g\vec{A}\left(\phi\phi^\dag+2vh\right)  
+ 2i\left(\vec{\partial}\phi\right)\phi^\dag \right]
\Biggl[-\frac{2}{3} 
-6\int_0^1 dx x(1-x)\ln\frac{\Delta(x,m_h^2)}{m_f^2}\Biggr]
\nonumber\\ &&\hspace{3cm}
+ f^4 \left( \phi\phi^\dag+ 4vh \right)
\phi\phi^\dag \Biggl[\frac{m_h^2}{4m_f^2}           
+ \frac{3}{2}\int_0^1 dx \ln\frac{\Delta(x,m_h^2)}{m_f^2}\Biggr] 
\Biggr\} \, ,
\end{eqnarray}
where $\phi=\Phi-v$ parameterizes the deviation of the
Higgs field from its vev.

\section{The Dirac Equation}

In this Appendix, we describe how we obtain the bound state energies of
the Dirac equation and the scattering phase shifts (and their Born
series) in the presence of a background potential.  These quantities
are required to compute the vacuum polarization energy in eq.~(\ref{Evac}).

The fermion field obeys the time-independent Dirac equation
\begin{equation}
H_D \Psi = \omega \Psi \, ,
\end{equation}
where 
\begin{equation}
H_D = -i\gamma^0\gamma^i\partial_i + \gamma^0 
\left[m_f + V(A,\Phi)\right] \, ,
\label{DiracEqn}
\end{equation}
and $V$ is given in eq.~(\ref{potential}).  In contrast to previous work, it is most convenient
to use the chiral representation of the Dirac matrices,
\begin{equation}
H_D \equiv \left( \begin{array}{cc} h_{11} & h_{12} \\ h_{21} & h_{22} 
\end{array} \right)  = \left( \begin{array}{cc} i\sigma_j\partial_j + 
g\sigma_jA_j & m_f(s+ip\tau_j\hat{x}_j ) \\ m_f(s-ip\tau_j\hat{x}_j ) & 
-i\sigma_j\partial_j \end{array} \right) \, .
\end{equation}    
The grand spin $\vec{G}$ is defined as the vector sum of isospin, spin and 
orbital angular momentum.  It commutes with $H_D$ as long as the fermion 
doublet is degenerate in mass and the background fields are in the spherical
ansatz.  We satisfy both conditions.  For a given grand spin quantum
number $G$ (we suppress the grand spin projection label $M$ throughout),
the Dirac spinor $\Psi_G$ has eight components and may be 
written in terms of generalized
spherical harmonic functions ${\mathcal Y}_{j,l} ({\hat{x}})$
with $j=G \pm \frac{1}{2}$ and $l=j \pm \frac{1}{2}$ as
\begin{equation}
\Psi_G (\vec{x}) = \left( \begin{array}{c} ig_1{\mathcal
Y}_{G+\frac{1}{2},G+1} + g_2{\mathcal Y}_{G+\frac{1}{2},G} +
g_3{\mathcal Y}_{G-\frac{1}{2},G} + ig_4{\mathcal
Y}_{G-\frac{1}{2},G-1} \\ i f_1{\mathcal Y}_{G+\frac{1}{2},G+1} +
f_2{\mathcal Y}_{G+\frac{1}{2},G} + f_3{\mathcal Y}_{G-\frac{1}{2},G}
+  if_4{\mathcal Y}_{G-\frac{1}{2},G-1} \end{array} \right) \, ,
\end{equation}
where $g_i$ and $f_i$ are radial functions and we have suppressed the
grand spin labels on them.  Note that in this chiral 
theory modes of different parity, {\it e.g.} $g_1$ and $g_2$
mix. The spherical harmonics are two-component spinors in both 
spin and isospin space.  The special case $\Psi_0$ is defined
only in terms of ${\mathcal Y}_{\frac{1}{2},1}$ and ${\mathcal
Y}_{\frac{1}{2},0}$ and does not contain $g_3, g_4, f_3, f_4$.

The matrix elements of operators like $\tau_j \hat{x}_j$ between the
spherical harmonics may be found in the literature \cite{Herbert}. 
We use them to write the Dirac equation (\ref{DiracEqn})
as a a set of eight coupled first-order linear differential equations 
in the radial functions, for fixed $G$. From these equations we
obtain the bound state
solutions ($| \omega |<m_f$) in  each grand spin channel using
shooting algorithms.  From  Levinson's theorem we determine the
number, $N_G^{\rm bound}$, of bound states to 
shoot for, using phase shifts,  $\delta_G (\omega)$, of the
scattering state solutions of the Dirac equation:
\begin{equation}
N^{\rm bound}_G = \frac{1}{\pi} \left( \delta_G(m_f) - 
\delta_G(\infty) + \delta_G(-m_f) - \delta_G(-\infty) \right) \, . 
\end{equation}
To construct these scattering state 
solutions we re-write the Dirac equation as a 
set of second-order differential equations in the radial functions.
Formally they read,
\begin{equation}
\left[ h_{12}h_{21} - h_{12}(h_{22}-\omega)h_{12}^{-1}(h_{11}-\omega) 
\right]\Psi_G^U = 0 \, ,
\label{upperdirac}
\end{equation}
where 
$$
\Psi_G^U=ig_1{\mathcal
Y}_{G+\frac{1}{2},G+1} + g_2{\mathcal Y}_{G+\frac{1}{2},G} +
g_3{\mathcal Y}_{G-\frac{1}{2},G} + ig_4{\mathcal
Y}_{G-\frac{1}{2},G-1}
$$
denotes the upper two-component spinor in $\Psi_G$.  
In the chiral representation of the Dirac matrices, we require 
$s^2(r)+p^2(r)>0$ so that $h_{12}$ is invertible. As 
mentioned in section V this is a restriction on our 
variational ans\"atze. Using the known matrix elements 
for the spin-isospin operators like $\tau_i\hat{x}_i$, we 
then project eq.~(\ref{upperdirac}) onto grand spin
channels and obtain the desired second order differential
equations. They may be written in the form,
\begin{equation}
\sum_{j=1}^{4} \left\{ D_G(r) + N_G(r)\frac{\partial}{\partial r} +
M_G(r) \right\}_{ij}g_j(r) = 0 \, ,
\end{equation}
with
\begin{eqnarray}
D_G(r) & = & \ID \left( \frac{\partial^2}{\partial r^2} +
\frac{2}{r}\frac{\partial}{\partial r} + k^2 \right) - \frac{1}{r^2}O_G \,
, \cr\cr
O_G & = & {\rm diag}\left((G+1)(G+2), G(G+1), G(G+1), (G-1)G\right)\, ,
\end{eqnarray}
and $k^2=\omega^2-m_f^2$. The matrices $N_G(r)$ and
$M_G(r)$ are given in terms of the functions 
$s(r),\ldots,\gamma(r)$ that specify the static background fields in the spherical ansatz, as in eq.~(\ref{SphericalAnsatz}). Their elements are rather 
lengthy and we refrain from explicitly displaying them here.
As $r\rightarrow\infty$, $N_G(r)\to0$ and $M_G(r)\to0$ and  
the differential equations decouple, as long as the potential goes 
to zero  sufficiently fast. 

We have a four-channel scattering problem.  We express the four wavefunctions 
and four boundary conditions in matrix form, ${\mathcal 
G}_{ij}(r)=g_i^{(j)}(r)$, where the linearly independent boundary 
conditions are labeled by $j=1,2,3,4$.  
We then write ${\mathcal G}(r)$ as a multiplicative modification of 
the matrix solution to the 
free differential equations, ${\mathcal G}(r)\equiv F(r)\cdot H(kr)$, 
where $H(x) = \mbox{diag}(h_{G+1}^{(1)}(x), h_{G}^{(1)}(x), 
h_{G}^{(1)}(x), h_{G-1}^{(1)}(x) )$ with
$h_{\ell}^{(1)}(x)$ denoting spherical 
Hankel functions of the first kind such that $D_G(r)\cdot H(kr)=0$. 
(The $4\times4$ matrices $F$ and $H$ depend on the 
grand spin quantum number $G$. For convenience we omit that
label from now on.)
Imposing the boundary conditions 
$F(r\rightarrow\infty)=0$ and 
$F'(r\rightarrow\infty)=0$, it is clear that the $i^{th}$ row of 
${\mathcal G}$ describes an outgoing spherical wave in the $i^{th}$ 
channel.  Similarly, ${\mathcal G}^{*}$ describes incoming spherical 
waves.  The scattering wavefunction can be written as 
\begin{equation}
{\mathcal G}_{\rm sc}(r) = -{\mathcal G}^{*}(r) + {\mathcal G}(r)S(k) 
\, ,
\end{equation}
and requiring this to be regular at the origin gives the scattering matrix
\begin{equation}
S(k) = \lim_{r\rightarrow 0}H^{-1}(kr)F^{-1}(r)F^{*}(r)H^{*}(kr) 
\, .
\end{equation}
We are interested in the sum of the eigenphase shifts in a given grand spin channel,
\begin{equation}
\delta(k) = \frac{1}{2i}\Tr \ln S(k) = 
\frac{1}{2i}\lim_{r\rightarrow 0}\Tr\ln \left( F^{-1}(r)F^{*}(r) 
\right) \, .
\end{equation}
An efficient way to avoid any ambiguities in 
additive contributions of multiples of $\pi$ in $\delta(k)$ is 
to define
\begin{equation}
\delta(k,r) = \frac{1}{2i}\Tr\ln \left( F^{-1}(r)F^{*}(r) \right) \, ,
\end{equation}
with $\delta(k)=\delta(k,0)$.
We then integrate
\begin{equation}
\frac{\partial \delta (k,r)}{\partial r} = - \Im\Tr\left( F'F^{-1} 
\right) 
\end{equation}
along with $F(k,r)$ from infinity to 0 with the boundary condition 
$\lim_{r\rightarrow\infty}\delta(k,r)=0$ to obtain $\delta(k)$ as a 
smooth function of $k$.  The differential equation for the matrix $F(k,r)$,
\begin{equation}
0 = F'' + \frac{2}{r}F' + 2F'L'+\frac{1}{r^2}[F,O]+N(F'+FL')+MF \, ,
\end{equation}
is obtained from  
\begin{equation}
\left[ \left\{ D(r)+ N(r)\frac{\partial}{\partial r} + M(r) \right\} 
{\mathcal G}(r) \right] H^{-1}(kr) = 0 \, ,
\end{equation}
where $L(kr)\equiv \ln H(kr)$ and primes denote
derivatives with respect to the radial coordinate.  
The components of $L'(kr)$ can be 
expressed as simple rational functions, which avoids any numerical 
instability that would be caused by the oscillating Hankel functions.

To construct the Born series for $\delta(k)$, we introduce 
$F^{(n)}(k,r)$ where $n$ labels the order in the background fields in an expansion around the \classicalVacuum configuration with $f^{(0)}(r) = 0$ in eq.~(\ref{Vacua1D}).  We obtain the corresponding differential equations
\begin{eqnarray}
0 & = & {F^{(1)}}'' + \frac{2}{r}{F^{(1)}}' + 
2{F^{(1)}}'L'+\frac{1}{r^2}[F^{(1)},O]+N^{(1)}L'+M^{(1)} \, , \nonumber 
\\
0 & = & {F^{(2)}}'' + \frac{2}{r}{F^{(2)}}' +
2{F^{(2)}}'L'+\frac{1}{r^2}[F^{(2)},O]+N^{(1)}\left({F^{(1)}}' +
F^{(1)} L'\right)  
\nonumber \\ && \hspace{1cm}
+N^{(2)}L' + M^{(1)}F^{(1)} + M^{(2)} \, ,
\end{eqnarray}
where the matrices $N^{(i)}$ and $M^{(i)}$ are obtained from 
$N$ and $M$ by expanding to order $i$ in the deviation of the 
background fields from the above described \classicalVacuum configuration.
We integrate these differential equations with the boundary
conditions $F^{(i)}(k,\infty)=0$ and $F^{(i)\prime}(k,\infty)=0$ and
obtain
\begin{eqnarray}
\delta^{(1)}(k)&=&-\Im \tr \left(F^{(1)}(k,0)\right)\,,
\nonumber \\ 
\delta^{(2)}(k)&=&-\Im \tr \left(F^{(2)}(k,0)-\frac{1}{2}
F^{(1)}(k,0)^2\right)\,.
\end{eqnarray}
We eliminate the quadratic divergence 
from the vacuum polarization energy by subtracting these from $\delta(k)$ and adding them back in as 
renormalized first and second order Feynman diagrams.  

There still remains the logarithmic divergence whose 
elimination would require third and fourth order Born subtractions.
These become considerably more complicated, so instead we use the 
limiting function approach as described in \cite{DecouplingNoGauge}.  The 
idea is to subtract only the local contributions to the third and fourth 
Born approximants to the phase shift by identifying them with the divergent 
contributions to the third and fourth order Feynman diagrams.  
To this end we formally manipulate these divergent 
Feynman diagrams. To extract the local contributions we 
set the external momenta to zero and then integrate over
the energy and the two spatial angles of the loop momenta, 
$k^\mu$, such that
a (regularized) integral over $k=|\vec{k}|$ is left. We 
write its integrand in the form as in eq.~(\ref{Evac}),
$$
\frac{1}{2\pi}\sqrt{k^2+m_f^2}\,\frac{d\delta_{\rm lim}(k)}{dk}
$$
where
\begin{eqnarray}
\delta_{\rm lim}(k) & = & \frac{1}{8\pi}\left(\frac{k}{k^2+m_f^2} + 
\frac{1}{m_f}\arctan\frac{m_f}{k} \right)
 \nonumber \\ && \hspace{0.5cm} \times
 \int d^3x\, \tr 
\Biggl\{\frac{g^3}{6}\left(4i\partial_iA_j+g[A_i,A_j]\right)[A_i,A_j] 
-m_f^4 \left(\phi\phi^\dagger + 4vh \right)\phi\phi^\dagger
\nonumber\\ && \hspace{3cm} 
+ m_f^2 g\vec{A}\cdot
\left[g\vec{A}\left(\phi\phi^\dagger+2vh\right)  
+ 2i\vec{\partial}\phi\phi^\dag \right]\Biggr\}
\label{LimitingPhaseshift}
\end{eqnarray}
in the $A_0 = 0$ gauge and where $\phi=\Phi-v$ denotes the 
deviation of the Higgs field from its vev.  

Thus we numerically determine the bound state energies, the
phase shifts, their Born series, and the limiting function.  These are
all ingredients in the expression for the fermion vacuum polarization
energy in eq.~(\ref{Evac}).
\pagebreak
\\

\end{document}